\documentclass[twocolumn,superscriptaddress,showpacs,prl]{revtex4}
\usepackage{amsmath,bm,amsfonts}
\usepackage[dvipdfm]{graphicx}
\def\ba{{\bm a}}
\def\bc{{\bm c}}
\def\bz{{\bm z}}
\def\bA{{\bm A}} \def\bC{{\bm C}}\def\bD{{\bm D}}
\def\bL{{\bm L}}\def\bP{{\bm P}}
 \def\bS{{\bm S}} \def\bT{{\bm T}}
\def\bZ{{\bm Z}}

\def\tq{\tilde{q}}
\def\tn{\tilde{n}}
\def\tc{\tilde{c}}
\def\tbc{\tilde{{\bm c}}}
\def\tk{\tilde{k}}
\def\tm{\tilde{m}}

\def\ga{\alpha}
\def\gvf{\varphi}

\def\GCD{\mathrm{GCD}}

\begin{document}
\title{Symmetry of arbitrary layer rolled-up nanotubes}
\author{M. Damnjanovi\'c} \author{B. Nikoli\'c} \author{I. Milo\v sevi\'c}
\email[]{ivag@afrodita.rcub.bg.ac.yu}
 \homepage[]{http//:nanolab.ff.bg.ac.yu}
 \affiliation {Faculty of Physics,
University of Belgrade, P.O. Box 368, 11001 Belgrade, Serbia}
\date{\today}
\begin{abstract}
Rolling up of a layer with arbitrary lattice gives nanotube with
very reach symmetry, described by a line group which is found for
arbitrary diperiodic group of the layer and chiral vector. Helical
axis and pure rotations are always present, while the mirror and
glide planes appear only for specific chiral vectors in rhombic and
rectangular lattices. Nanotubes are not translationally periodical
unless layer cell satisfies very specific conditions. Physical
consequences, including incommensurability of carbon nanotubes, are
discussed.
\end{abstract}

\maketitle

\section{Introduction}

Large line group symmetry of carbon nanotubes (NTs) is
substantial~\cite{WHITE,YCSYM} in predicting their unique
properties~\cite{ROTKIN,RotkinBook}. The underlying 2D hexagonal
lattices enabled to apply essentially the same symmetry to several
inorganic NTs~\cite{MS2}. However, besides recent result on the
rectangular lattices~\cite{LGRECTANGULAR}, the symmetry of NTs
related to the other kinds of 2D lattices has never been considered,
despite rapidly increasing number of the reported types:
BC$_2$N~\cite{MIY94}, ternary borides~\cite{DEZA00}, carbon
pentaheptides and Haeckelites~\cite{C57}, ZnO
nanorings~\cite{KONG04}, etc.

Here we fill in this gap presenting quite general result: full
symmetry of the rolled-up arbitrary layer is found. NT may be
without translational periodicity, but its symmetry is always
described by a line group. As discussed, this has many far reaching
physical consequences, including reconsideration of carbon NTs'
symmetry.

\section{Line groups}\label{SLG}

In contrast to 2D and 3D crystals, quasi-1D ones are not subdued to
the crystallographic restrictions on the rotational axis, while in
some cases helical ordering substitutes translational periodicity.
Consequently, number of different symmetry groups, {\em line
groups}, of quasi-1D systems is infinite, in contrast to 80
diperiodic and 273 space groups. Only 75 line groups are subgroups
of the latter; they are known as rod groups~\cite{KOP03}.

Line groups are classified within thirteen families
(Tab.~\ref{TLGfold}). Each group is a product $\bL=\bZ\bP_n$ of an
axial point group $\bP_n$ and an infinite cyclic group $\bZ$. Thus
$\bP_n$ is one of $\bC_{n}$, $\bS_{2n}$, $\bC_{nh}$, $\bD_{nv}$,
$\bC_{nv}$, $\bD_{nd}$, $\bD_{nh}$, where $n=1,2,\dots$ is the order
of its principle axis ($z$-axis, by convention). Group of the {\em
generalized translations} $\bZ$ is either screw axis $\bT_Q(f)$ or a
glide plane $\bT'(f)$, generated by $(C_Q|f)$ (Koster-Seitz symbol),
i.e. rotation for $2\pi/Q$ around the $z$-axis followed by
translation for $f$ along the same axis ($Q\geq1$ is a real number),
and $(\sigma_v|f)$, respectively. While pure translation (for $2f$)
pertains to $\bT'(f)$, group $\bT_{Q}(f)=\bT^{r}_{q}(f)$ contains
pure translation (for $fq$) only for rational $Q=q/r$ ($r$ and $q$
integers).

\begin{figure}[hbt]\centering
 \includegraphics[width=8cm]{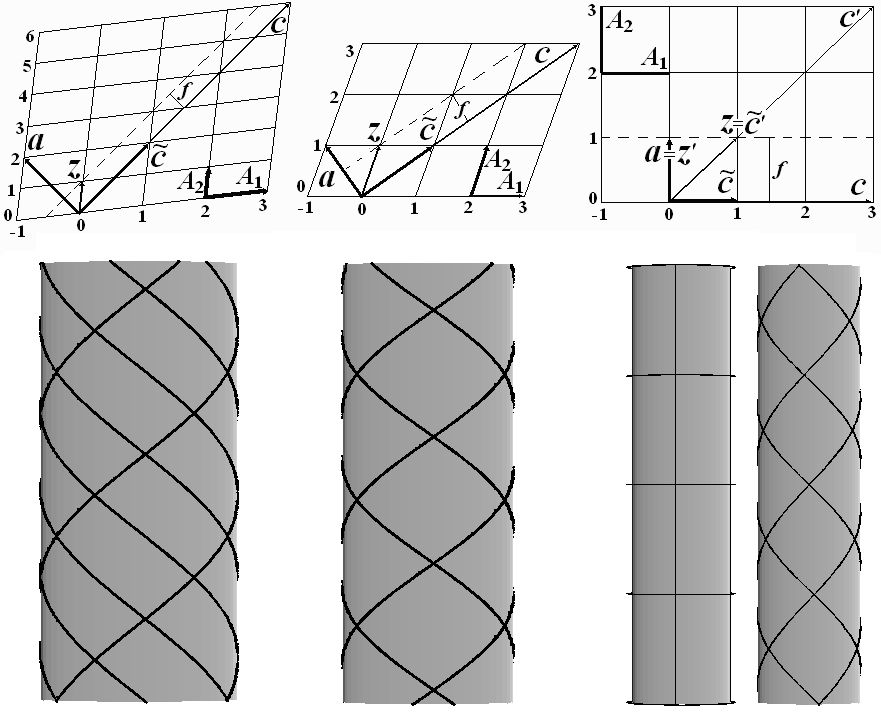}

  \caption{Top: Chiral vectors $\bc$, with corresponding
  $\ba$ (in the commensurate cases),
  $\bz$ (all $\bz_s$ are on the dashed line) and
  $f$. Left: $\bA_1=(12,\sqrt{2})$ and
  $\bA_2=(\tfrac{1}{\sqrt{2}},6)$ ($X=w=4$,
  $Y=J+X=24\sqrt{2}/73$, $x=0$, $y=1$), with $\bc=(3,6)$, $\bz=(0,1)$ and
  $\bL^{(1)}=\bT^1_{12}(3\sqrt{2}-\tfrac12)\bC_3$
  Middle: rhombic layer with $\ga=70^\circ$, $\bc=(3,3)$, $\bz=(0,1)$
  and $\bL^{(1)}=\bT^1_6(a/2\approx0.57A_1)\bC_3$. Right: rectangular layer with $A_1/A_2=\pi/3$,
  $\bc=(3,0)$, $\bz=(1,1)$, $\bL^{(1)}=\bT^1_3(a=A_2)\bC_3$ and
  $\bc'=(3,3)$, $\bz'=(0,1)$, $\bL^{(1)'}=\bT^1_{\tfrac{9+\pi^2}{3}}(\tfrac{3\pi}{\sqrt{9+\pi^2}}A_1)\bC_3$. Bottom: corresponding
  nanotubes.}\label{Flayer}
\end{figure}

All the combinations $(C_Q|f)^tC^s_n$ of rotations around the
principle axis and translations form roto-helical subgroup
$\bL^{(1)}$ (first family line group) of $\bL$. In the course of
rolling it emerges from the 2D lattice translations. For the first
family line groups $\bL^{(1)}=\bL$; for the groups from the 2-8
families $\bL^{(1)}$ is a halving subgroup, while for the families
9-13 $\bL^{(1)}$ is a quarter subgroup. Consequently, when
$\bL^{(1)}$ is known, to build up the whole line group it remains to
find if there are additional generators (mirror/glide plane and/or
$U$-axis) allowed both by the symmetry of the elementary cell of the
specific layer and chiral vector.

Note that $Q$ is not unique, and therefore the
convention~\cite{LGRECTANGULAR} is introduced: $Q$ is the greatest
finite among $Q_s=Qn/(Qs+n)$; for the commensurate groups this $Q$
is written as $Q=q/r$ with $q$ being multiple of $n$, $q=\tq n$, and
$r$ is coprime to $\tq$, while the translational period is $a=\tq
f$. For example, the group combining pure translations $\bT(f=a)$
with $\bC_n$ is $\bT(a)\bC_n=\bT^1_n(a)\bC_n$.

Only the line groups from the first and fifth family may have
irrational $Q$, in which case they refer to the {\em incommensurate
systems}, i.e. helically ordered but with no translational
periodicity. The other families are characterized by $Q$ equal to
$n$ or $2n$, i.e. these groups are achiral $\bT(f)\bP_n$ (symmorphic
line groups), $\bT_{2n}(f=a/2)\bP_n$ or $\bT'(f=a/2)\bP_n$.

\section{Roto-helical transformations}\label{SLGHEL}

We consider 2D lattice, with a basis $\bA_1$ and $\bA_2$ ($A_1\geq A_2$) at the
angle $\ga\in(0,\pi/2]$ and define dimensionless parameters $X$ and $Y$:
\begin{equation}\label{EX}
  X=\frac{A^2_1}{A^2_2}\geq1,\quad Y=\frac{A_1}{A_2}\cos\ga\geq0.
\end{equation}
NT $(n_1,n_2)$ is obtained by folding the layer in a way that the
{\em chiral vector} $\bc=(n_1,n_2)=n_1\bA_1+n_2\bA_2$ becomes
circumference of the tube (Fig.~\ref{Flayer}). Alternatively, NT is
defined by length $c$ (giving the tube's diameter $D=c/\pi$) and
slope $\theta$ (called {\em chiral angle}) of $\bc$:
\begin{equation}\label{Etc}
c=A_2\sqrt{n^2_1X+n^2_2+2n_1n_2Y},\quad\sin\theta=n_2A_2/c.
\end{equation}
It is enough to consider NTs with $n_2\geq0$, i.e.
$0\leq\theta<\pi$, as the nanotube $(-n_1,-n_2)$ is same with
$(n_1,n_2)$.

The translations of the layer become roto-helical operations on the
tube, i.e. two-dimensional translational group is folded into
$\bL^{(1)}=\bT_Q(f)\bC_n$. Simple geometry and some number theory
suffice to find the parameters $Q$, $n$ and $f$~\cite{RemTilt}:
\begin{subequations}\label{Etubelg}
\begin{eqnarray}
\label{Etuben}n&=&\GCD(n_1,n_2),\\
\label{Etubef}f&=&A_1\frac{\sin\ga}{\sqrt{\tn^2_1X+\tn^2_2+2\tn_1\tn_2Y}},\\
\label{EtubeQ}
  Q&=&n\frac{\tn^2_1X+\tn^2_2+2\tn_1\tn_2Y}
       {\tn_1z_1X+\tn_2z_2+(\tn_1z_2+\tn_2z_1)Y}.
\end{eqnarray}\end{subequations}
Here, $\bz=(z_1,z_2)$ is the closest to the line perpendicular to
$\bc$ (but not on this line, Fig.~\ref{Flayer}) lattice vector from
the series
\begin{eqnarray}\label{Elgzsol}
 \bz_s&=&(z_{1s},z_{2s})=\bz_0+s(\tn_1,\tn_2),\quad s=0,\pm1,\dots\\
\bz_0&=&
  \begin{cases}
    (0,1), & \text{if $\bc=(n,0)$}, \\
    (-1,0), & \text{if $\bc=(0,n)$}, \\
    (\tn_2^{\gvf(\tn_1)-1},\frac{\tn_2^{\gvf(\tn_1)}-1}{\tn_1}), &
    \text{otherwise.}
  \end{cases}
\end{eqnarray}
The Euler function $\gvf(x)$ gives the number of co-primes with $x$
which are less than $x$.

In conclusion, the roto-helical part $\bL^{(1)}$ of the NT symmetry
generates the whole tube from a single 2D unit cell. The symmetry
parameters $f$ and $\tilde{Q}=Q/n$ depend only on the reduced chiral
vector $\tbc=\bc/n=\tn_1\bA_1+\tn_2\bA_2$, and thus they are the
same for the ray of the NTs $n(\tn_1,\tn_2)$ differing by the order
$n$ of the principle axis.

\subsection{Commensurability}\label{SCOM}

Instead of analyzing when \eqref{EtubeQ} gives rational $Q$,
commensurability condition of NT is found by a direct check of
existence of NT's translational period $a$. Obviously, if exists,
$a$ is the length of the minimal lattice vector
$\ba=a_1\bA_1+a_2\bA_2$ orthogonal onto the chiral vector. So we
look for solvability (in coprime integers $a_i$) of
\begin{equation}\label{ElaAcond}
    \tbc\cdot\ba=a_2\tn_2+a_1 \tn_1 X+(a_2 \tn_1+a_1\tn_2)Y=0.
\end{equation}
When solvable, the period of the tube is
\begin{equation}\label{EcomA1A2}
a=A_2\sqrt{a^2_1X+a^2_2+2a_1a_2Y}.
\end{equation}
Further, as $q$ is the number of 2D lattice unit cells within a
translational period of a NT, the surface areas equality,
$qA_1A_2\sin\ga=c a$, gives:
\begin{equation}\label{Eqcomm}
  q=n\frac{\tc a}{A^2_2\sqrt{X-Y^2}}.
\end{equation}
Finally, $r$ is to be found from \eqref{EtubeQ}.

To discuss solvability of \eqref{ElaAcond}, we note that only $X$
and $Y$ may be irrational. As the reals are an infinite dimensional
vector space over the rational numbers, for solvability in rational
$a_i$ (then also integral solutions exist) it is necessary that 1,
$X$ and $Y$ are rationally dependent: either both $X$ and $Y$ are
rational, or there are rational $w$, $x$ and $y$ (with $x\neq y$ as
$X>Y$) and irrational $J$, such that $X=w+xJ$ and $Y=w+yJ$.

For both $X$ and $Y$ rational, Eq.~\eqref{ElaAcond} becomes a
(rational) proportion between $a_1$ and $a_2$. All the NTs
$(n_1,n_2)$ are commensurate with~\cite{RemTilt}:
\begin{eqnarray}\label{Eqrac}
 q&=&n\frac{2\tn_1\tn_2\underline{X}\overline{Y}+\tn_1^2\overline{X}\underline{Y}+\tn_2^2\underline{X}\underline{Y}}
    {\GCD(\tn_1\underline{X}\overline{Y}+\tn_2\underline{X}\underline{Y},\tn_2\underline{X}\overline{Y}+\tn_1\overline{X}\underline{Y})},\\
\label{Earac}
 \ba&=&\frac{\left(\tn_1\underline{X}\overline{Y}+\tn_2\underline{X}\underline{Y},-\tn_2\underline{X}\overline{Y}-\tn_1\overline{X}\underline{Y}\right)}
  {\GCD(\tn_1\underline{X}\overline{Y}+\tn_2\underline{X}\underline{Y},\tn_2\underline{X}\overline{Y}+\tn_1\overline{X}\underline{Y})}.
\end{eqnarray}

In the other case, rational and irrational parts of \eqref{ElaAcond}
are system of two homogeneous equations in $a_i$, solvable when its
determinant vanishes:
\begin{equation}\label{Edet}
   \tn_1^2w(y-x)-\tn_1\tn_2x-\tn^2_2 y =0.
\end{equation}
This constraint on $\tn_1$ and $\tn_2$ singles out a subset of the
chiral vectors yielding commensurate NTs. Note that solutions may
not exist (rationals are not algebraically closed). When exist, all
the chiral vectors $n\tbc$ ($n=1,2,\dots$) give commensurate NTs
with period $\ba$, being fully determined by the reduced chiral
vector $\tbc$. Besides, interchanging roles of $\tbc$ and $\ba$, we
get NTs with period $\tbc$, orthogonal onto chiral vectors $n\ba$.
Hence, in such a lattice commensurate NTs lie on two perpendicular
lines.

For $\tn_2\neq0$, the constraint \eqref{Edet} becomes
\begin{equation}\label{Esolution}
 \frac{\tn_1}{\tn_2}=\frac{x\pm\sqrt{x^2-4wxy+4wy^2}}{2w(y-x)}=\nu_\pm,
\end{equation}
and  $\tn_1/\tn_2$ is rational only if $\sqrt{x^2-4wxy+4wy^2}$ is.
The commensurate NTs are $\bc^\pm=n\tbc^\pm$ with mutually
orthogonal
$\tbc^\pm=\ba^\mp=(\overline{\nu}_\pm,\underline{\nu}_\pm)$, giving
$q=n(\underline{\nu}_+\overline{\nu}_--\underline{\nu}_-\overline{\nu}_+)$
and
 $a^\pm=A_2\sqrt{\underline{\nu}^2_\mp + \overline{\nu}^2_\mp X +
 2\underline{\nu}_\mp\overline{\nu}_\mp Y}$.

The case $\tn_2=0$ appears if and only if $w=0$, meaning $Y=yX$
(i.e. $J=X$ and $x=1$). Then the ray orthogonal onto $\tbc^+=(1,0)$
is obtained as $\tbc^-=(\overline{y},\underline{y})$. The periods
are $a^+=A_1\overline{y}|\tan\ga|$ and $a^-=A_1$, while $q=n
\underline{y}$.

\section{Additional symmetries}\label{Stotlg}

Apart from the translational symmetry, a 2D lattice has a rotational
$\bC_2$ symmetry which is generated by the rotation for $\pi$ around
the axis perpendicular to layer. In addition, rhombic and
rectangular lattices have vertical mirror and glide planes  and
also, in rhombic rectangular and hexagonal lattices the order of the
rotational axis is four and six, respectively. However, atomic
arrangements within the lattice unit cell may reduce the symmetry
group to one of 80 diperiodic groups~\cite{KOP03}. The additional
symmetries, preserved  after rolling up a layer into a NT may
appear: two-fold rotational axis, mirror and glide planes. When
combined with the roto-helical group $\bL^{(1)}$ given by
\eqref{Etubelg}, they yield a line group which belongs to one of the
remaining twelve families.

Rotation $C_2$ of a layer becomes horizontal two-fold axis, the
$U$-axis, of the tube. Thus, whenever order of the principle axis of
the layer is two, four, or six, symmetry of the NT is the fifth
family line group $\bT_Q(f)\bD_n$ at least. Note that the higher
order rotational symmetries of the layer do not give rise to the
symmetry of NTs.

Vertical mirror (glide) plane is preserved in the NT only if the
chiral vector is perpendicular onto it. When $\bc$ is parallel to
the plane, NT gets horizontal mirror (roto-reflectional) plane. All
these transformations can be combined (Tab.~\ref{TLGfold}) only with
the roto-helical groups $\bT(a)\bC_n$ or $\bT^1_{2n}(a)\bC_n$ (i.e.
$\tq=1,2$) of the achiral NTs.

Firstly, we consider rectangular lattices, $\ga=\pi/2$ (i.e. $Y=0$).
For irrational $X=J$, we have $w=y=0$, (then $\underline{y}=1$) and
$x=1$, yielding $\tc^+=(1,0)$ and $\tc^-=(0,1)$, with $\tq=1$, i.e.
the helical factor reduces to the pure translational group. For $X$
rational, from Eq.~\eqref{Eqrac} we get
$\tq=(\tn^2_1\overline{X}+\tn^2_2\underline{X})/\GCD(\tn_1,\underline{X})\GCD(\tn_2,\overline{X})$.
Thus, for $X\neq1$, the same result as for $X$ irrational is
achieved, while in the case $X=1$ (square lattice) additionally
$\tq=2$ is obtained for $\tc^\pm=(\pm1,1)$.

\squeezetable
\begin{table}[htb]\centering \caption{\label{TLGfold}Rolling-up correspondence of the line
and diperiodic groups. For each family (column 1) of the linegroups
all the different factorizations, roto-helical subgroup $\bL^{(1)}$
and the isogonal point group $\bP^I_q$ are given in the columns 2, 3
and 4 (for irrational $Q$ in the families 1 and 5, $q$ is infinite;
$\bT'_{d}$ is glide plane bisecting vertical mirror planes or
$U$-axes of $\bP$). The corresponding diperiodic groups enumerated
according to Ref.~\onlinecite{KOP03} follow: for arbitrary chiral
vector rolling gives either the first or fifth family line group;
only for special chiral vector(s) $\bm{a}=(n,0)$, $\bm{b}=(0,n)$,
$\bm{c}\in\{(n,0),(0,n)\}$, $\bm{d}=(n,n)$, $\bm{e}=(-n,n)$,
$\bm{f}\in\{(n,n), (-n,n)\}$, $\bm{g}\in\{(n,0), (0,n), (-n,n)\}$,
$\bm{h}\in\{(n,n), (-n,n), (-2n,n)\}$, $\bm{i}\in\{(n,0),
(0,n),(-n,n),(n,n), (-n,2n), (-2n,n)\}$ the underlined groups
(repeated after the corresonding vectors) give other line group
families below.}

\begin{tabular} {lllll}\toprule
F&Factorizations&$\bL^{(1)}$&$\bP^I_q$ & DG  \\
 \colrule
 1&$\bT_Q\bC_n$&$\bT_Q\bC_n$&$\bC_q$& 1,2,4,5,8,9,10,\underline{11},\underline{12},\underline{13},\underline{14},\\
  &&&& \underline{15},\underline{16},\underline{17},\underline{18},\underline{27},\underline{28},\underline{29},\underline{30},\underline{31}, \\
  &&&& \underline{32},\underline{33},\underline{34},\underline{35},\underline{36},65,66,67,68, \\
  &&&& \underline{69},\underline{70},\underline{71},\underline{72},74,\underline{78},\underline{79} \\ 
 5&$\bT_Q\bD_n$&$\bT_Q\bC_n$&$\bD_q$&3,6,7,\underline{19},20,21,22,\underline{23},\underline{24},\underline{25},\\
  &&&& \underline{26},\underline{37},\underline{38},\underline{39},\underline{40},\underline{41},\underline{42},\underline{43},\underline{44},\\
  &&&& \underline{45},\underline{46},\underline{47},\underline{48},49,50,51,52,53,\\
  &&&& 54,\underline{55},\underline{56},\underline{57},\underline{58},\underline{59},\underline{60},\underline{61},\underline{62},\\
  &&&& \underline{63},\underline{64},73,75,76,\underline{77},\underline{80} \\ 
 2&$\bT{\bS}_{2n}$&$\bT\bC_n$&$\bS_{2n}$& $\bm{a}$:17,33,34; $\bm{b}$:12,16,29,30\\ 
 3&$\bT\bC_{nh}$&$\bT\bC_n$&$\bC_{nh}$&$\bm{a}$:11,14,15,27,31,32;$\bm{b}$:28\\ 
 4&$\bT^1_{2n}\bC_{nh},\bT^1_{2n}\bS_{2n}$&$\bT^1_{2n}\bC_n$&$\bC_{2nh}$&$\bm{e}$:13,18,35; $\bm{d}$:36; \\
 & & & & $\bm{h}$:69,72,78; $\bm{g}$:70,71,79  \\ 
 6&$\bT\bC_{nv},\bT'_d\bC_{nv}$&$\bT\bC_n$&$\bC_{nv}$&$\bm{a}$:28; $\bm{b}$:11,14,15,27,31,32\\ 
 7&$\bC_{n}\bT'$&$\bT\bC_n$&$\bC_{nv}$& $\bm{a}$:12,16,29,30; $\bm{b}$:17,33,34\\ 
 8&$\bT^1_{2n}\bC_{nv},\bT'_d\bC_{nv}$&$\bT^1_{2n}\bC_n$&$\bC_{2nv}$&$\bm{e}$:36; $\bm{d}$:13,18,35; \\
 & & & & $\bm{h}$:70,71,79; $\bm{g}$:69,72,78  \\ 
 9&$\bT\bD_{nd},\bT'\bD_{nd}$&$\bT\bC_n$&$\bD_{nd}$& $\bm{a}$:42,45; $\bm{b}$:24,38,40\\ 
 10&$\bT'\bS_{2n}=\bT'_d\bD_n$&$\bT\bC_n$&$\bD_{nd}$& $\bm{c}$:25,39,43,44,56,60,62,63\\ 
 11&$\bT\bD_{nh},\bT'\bD_{nh}$&$\bT\bC_n$&$\bD_{nh}$& $\bm{c}$:23,37,41,46,55,59,61,64\\ 
 12&$\bT'\bC_{nh},\bT'\bD_n$&$\bT\bC_n$&$\bD_{nh}$&$\bm{a}$:24,38,40; $\bm{b}$:42,45\\ 
 13&$\bT^1_{2n}\bD_{nh},\bT^1_{2n}\bD_{nd},$&$\bT^1_{2n}\bC_n$&$\bD_{2nh}$& $\bm{f}$:26,47,48,55,56,57,58,\\
  &$\bT'_d\bD_{nh}, \bT'_d\bD_{nd}$ & & & $\bm{f}$:61,62,63,64; $\bm{i}$:77,80\\ \botrule
\end{tabular}\end{table}

Secondly, in the case of rhombic lattices $A_1=A_2$ ($X=1$) for $Y$
irrational, taking $J=Y-1$, we have $w=y=1$ and $x=0$ (then
$\underline{x}=1$), yielding $\tc^\pm=(\pm1,1)$ with $\tq=2$, i.e.
$\bL^{(1)}=\bT^1_{2n}(a/2)\bC_n$. For rational $Y$, from
\eqref{Eqrac} we get
$\tq=(2\tn_1\tn_2\overline{Y}+(\tn^2_1+\tn^2_2)\underline{Y})
/\GCD(\tn_1\overline{Y}+\tn_2\underline{Y},\tn_1\underline{Y}+\tn_2\overline{Y})$,
allowing the same $\tc^\pm$ as for $Y$ irrational. Only for $Y=0$
and $Y=1/2$ the additional pair, $\tbc^+=(1,0)$ and $\tbc^-=(0,1)$,
appear, giving $\tq=1$ in a case of the square lattice, and again,
$\tq=2$ for the hexagonal lattice.

Finally, additional symmetries of the layer reduce the number of the
different nanotubes. For the layers with the principle axis order
$n=1,2,3,4,6$, the effective interval of the chiral angle is
$[0,2\pi/n')$, where $n'=\mathrm{LCM}(2,n)=2,2,6,4,6$, respectively.
Further, vertical mirror plane of the layer intertwins the chiral
vectors of the optically isomeric tubes, enabling to halve this
range to $[0,\pi/n']$. However, if there is not such a plane, the
optical isomer of tube $(n_1,n_2)$ is obtained from the layer
reflected in the mirror plane (perpendicular to $\bA_1$) again as
the tube $(n_1,n_2)$.

\section{Discussion}\label{SDIS}
Full symmetries of NTs rolled up from arbitrary 2D layers for any
chiral vector are described by line groups. Depending on 2D lattice
and direction of chiral vector, NT may be incommensurate (line group
from the first or fifth family) or commensurate. Without optical
isomers (achiral) are NTs with pure translational or zig-zag helical
factors; they are always commensurate, and obtained for special
chiral vectors from the rectangular and rhombic 2D lattice
respectively, allowing mirror/glide planes. Presented results agree
with the known ones in the cases of hexagonal~\cite{WHITE,YCSYM} and
rectangular~\cite{LGRECTANGULAR} lattices.

The conserved quantum numbers related to the roto-helical symmetries
of NTs are quasi-momenta~\cite{WHITE,YCSYM}: helical, $\tk$, from
the helical Brillouin zone (HBZ) $(-\pi/f,\pi/f]$, and remaining
(non-helical part) angular $\tm$, taking integer values from the
interval $(-n/2,n/2]$. When $U$-axis, vertical or horizontal
mirror/glide plane is a symmetry, the corresponding parity ($\Pi^U$,
$\Pi^v$ and $\Pi^h$, taking values $+1$ and $-1$ for even and odd
states) is conserved. More conventional quantum numbers for
commensurate nanotubes are linear and total angular quasi-momenta,
$k$ varying within Brillouin zone (BZ) $(-\pi/a,\pi/a]$) and $m$
(integers from $(-q/2,q/2]$), where $q$ is the order of the
principle axis of the isogonal point group. However, $m$ is not
conserved in the Umklapp processes.

These quantum numbers assigning energy bands $E^\Pi_{\tm}(\tk)$ (or
$E^\Pi_m(k)$) of (quasi)particle spectra, correspond to irreducible
representations of the NT's line group; the dimension of
representation is degeneracy of the band. Thus, for incommensurate
NTs, degeneracy is either one or two, while for the commensurate
also fourfold degeneracy is possible (families 9-13). In the
families 2-5 and 9-13, $z$-reversing elements ($U$-axis, horizontal
mirror/roto-reflectional plane) intertwine $\tk$ and $-\tk$ ($k$ and
$-k$), causing at least twofold band degeneracy in the interior of
the reduced HBZ $[0,\pi/f]$ (reduced BZ $[0,\pi/a]$). Only at its
boundaries 0 and $\pi/f$ (0 and $\pi/a$) this degeneracy is absent
(but simultaneous intertwining of $\pm\tm$ and $\pm m$ caused by
$U$-axis may preserve degeneracy for $\tm\neq0,n/2$ and $m\neq0,q/2$
in these points). Thus only these boundaries are special $\tk$- and
$k$-points, where the bands joining and van Hove singularities
appear. Let us mention that the NT's electronic subsystem is in the
nondegenerate state (excluding spin), as the Jahn-Teller theorem
hold for the line groups~\cite{IJT}.

The groups of the same family with same $n$ are isomorphic for any
$Q$, and change of this continual parameter does not diminish
symmetry. Accordingly, $Q$ should be varied in numerical relaxation.
For the most studied NTs, carbon NTs, the graphene lattice is
rhombic hexagonal, i.e. $X=1$ and $Y=\frac12$, giving commensurate
NTs of fifth family, besides achiral zig-zag and armchair tubes of
13th family~\cite{YCSYM}. However, for the chiral tubes, the
relaxation slightly changes helical axis, and there is no {\em a
priori} physical reason for commensurability. Incommensurability
affects the physical properties: quasi momentum fails to be
conserved, and only the selection rules of the helical quantum
numbers~\cite{WHITE,IJT} are applicable. This effect must be taken
into account, particularly when external fields or mechanical
influence~\cite{ROTKIN} like twisting is studied.




\begin{thebibliography}{99}
\bibitem{RemTilt}
Tilt quantities are divided by $n$. Over- and under-bar denote
numerator and denominator of rationals:
$x=\overline{x}/\underline{x}$.
\end{thebibliography}
\end{document}